\pdfoutput=1
\documentclass[conference]{IEEEtran}

\usepackage{cite}
\usepackage{flushend}
\usepackage{amsmath,amssymb,amsfonts}
\usepackage{multirow}
\usepackage{graphicx}
\graphicspath{ {./Images/} }
\usepackage{textcomp}
\usepackage[table]{xcolor}
\usepackage{booktabs}
\usepackage{tikz}
\usepackage{url}
\usetikzlibrary{positioning}
\def\BibTeX{{\rm B\kern-.05em{\sc i\kern-.025em b}\kern-.08em
    T\kern-.1667em\lower.7ex\hbox{E}\kern-.125emX}}

\usepackage{eso-pic}
\AddToShipoutPictureFG*{%
  \AtPageUpperLeft{%
    \put(\LenToUnit{\dimexpr(\paperwidth-\textwidth)/2\relax},\LenToUnit{-0.30in}){%
      \parbox[t]{\textwidth}{\centering\scriptsize
        Accepted for publication in the 2026 IEEE Cyber Awareness and Research Symposium (CARS).\\
        \copyright~2026 IEEE. Personal use of this material is permitted. Permission from IEEE must be obtained for all other uses, in any current or future media, including reprinting/republishing this material for advertising or promotional purposes, creating new collective works, for resale or redistribution to servers or lists, or reuse of any copyrighted component of this work in other works.}}}}

\begin{document}

\title{STRIATUM-CTF: A Protocol-Driven Agentic Framework for Autonomous CTF Solving}

\author{
	\IEEEauthorblockN{
    James Hugglestone\IEEEauthorrefmark{1},
    Samuel Jacob Chacko\IEEEauthorrefmark{1},
    Dawson Stoller,
    Ryan Schmidt,
    Xiuwen Liu
    }
	\IEEEauthorblockA{
    \textit{Department of Computer Science}, \textit{Florida State University}, Tallahassee, USA \\
    Email: jah21e@fsu.edu, sj21j@fsu.edu, dawson.stoller@aol.com, res22f@fsu.edu, xliu@fsu.edu
    }
    \IEEEauthorrefmark{1}These authors contributed equally.
}

\maketitle

\begin{abstract}
Autonomous cyber operations increasingly depend on agents that can chain reconnaissance, exploitation, and analysis across many tools while maintaining state over long task horizons. Large Language Models (LLMs) are promising orchestrators for this workflow, but naive agents that pipe model output directly into a shell are brittle: they emit malformed commands, hallucinate non-existent command-line options or packages, and lose track of earlier findings as the context window fills. We present \textbf{STRIATUM-CTF}, a modular agentic framework for Capture-the-Flag (CTF) solving that mediates between LLM reasoning and tool execution using the Model Context Protocol (MCP). Offensive-security tools are exposed as strictly typed, schema-validated functions; the protocol layer rejects out-of-schema tool calls before execution and returns structured observations that the agent uses to self-correct. We evaluate the system in two ways. In a controlled study of 180 runs across 15 CTF challenges under four levels of attached procedural documentation, the system succeeds on 77.8--86.7\% of attempts; notably, a schema-only configuration with no procedural documentation already reaches 77.8\%, and no condition differs significantly from another. In a single live, multi-team CTF at Florida State University, the system placed first against a field of more than 30 registered teams---a single, non-replicated event decided by a 10-point margin over the runner-up, and therefore a demonstration of feasibility rather than of superiority over human experts. Because all four ablation conditions share the same tool layer, our controlled study isolates documentation volume rather than the protocol layer itself; we accordingly present schema-mediated grounding as a design principle with qualitative support rather than as a measured effect. We will release our code, challenge set, per-run outcome data, and analysis scripts upon publication to support replication.
\end{abstract}


\section{Introduction}

Offensive security assessments, including penetration testing, red-teaming, and Capture-the-Flag (CTF) competitions, remain predominantly manual and labor-intensive. To compromise a target, an analyst enumerates attack surfaces, interprets verbose scanner output, selects exploitation primitives, and iteratively tunes payloads. While powerful tools exist for individual phases (e.g., Nmap, Ghidra, Hashcat), orchestrating them remains a cognitive bottleneck. Human operators act as the ``glue,'' manually moving data between tools, a process prone to error and cognitive biases such as confirmation bias~\cite{aggarwal2024discovering}.

Large Language Models (LLMs)~\cite{vaswani2017attention} can process unstructured text and generate code, making them well suited to automate the high-level reasoning that connects these tools. However, early autonomous security agents face architectural hurdles~\cite{hakim2025neuro}. Existing methods largely fall into two categories: ``copilots'' that require constant human prompting, and ``naive autonomous agents'' that pipe model output into a shell~\cite{happe2025surprising}. Models often suffer a \textit{reasoning--action gap}~\cite{stechly2025self}, correctly identifying a vulnerability (e.g., SQL injection) yet failing to emit a syntactically valid payload, frequently inventing non-existent command-line options or software packages (the supply-chain risk known as \textit{slopsquatting}~\cite{park2025slopsquatting}). On long-horizon tasks, ``context drift'' causes the agent to lose track of early reconnaissance data~\cite{liu2024lost,sood2025paradigm}, producing loops of redundant, failed attempts.

To address these limitations, we introduce \textbf{STRIATUM-CTF},\footnote{From ``Search-based Test-time Reasoning Inference Agent for Tactical Utility Maximization''; we use STRIATUM simply as the system's name.} an agentic framework that mediates between LLM reasoning and tool execution using the Model Context Protocol (MCP)~\cite{mcp_docs}. The LLM never executes directly: it emits structured tool calls that the protocol layer validates against a schema and runs, returning typed results rather than free-form text. Internally, the agent is equipped with offensive primitives including symbolic execution via Angr~\cite{shoshitaishvili2016sok}, static analysis with Ghidra~\cite{ghidra2019}, and runtime debugging through GDB. Externally, the environment exposes reconnaissance and exploitation utilities such as Nmap, Nuclei, FFUF, and TLSX. Given a high-level objective, STRIATUM-CTF formulates a multi-step plan, invokes tools via structured MCP calls, parses error streams to self-correct, and iterates until the objective is met or the search budget is exhausted.

This paper makes three contributions:
1) \textbf{A protocol-driven agentic architecture:} we use MCP to standardize LLM interaction with offensive-security tools, rejecting out-of-schema (malformed) tool invocations before execution and enabling error recovery through structured feedback. We advance this as a \emph{design principle} supported by qualitative evidence, not as a measured effect: the experiments reported here hold the tool layer fixed across all conditions and therefore cannot attribute outcomes to the protocol layer (Sections~\ref{sec:limits} and~\ref{sec:future});
2) \textbf{Integration of specialized analysis primitives:} we expose symbolic execution (Angr) and dynamic debugging (GDB) as in-context tools for binary analysis and exploit generation; and
3) \textbf{Two-pronged evaluation:} a 180-run controlled ablation, reported per-challenge, showing performance is robust to a roughly $60\times$ reduction in attached documentation tokens, and a first-place finish in a live, multi-team CTF against human teams, reported with explicit statistical scope.
A companion study~\cite{chacko2026skills} re-analyzes the same 180 runs as an evaluation of Agent Skills---structured procedural-knowledge packages attached to an agent's context---and reaches a complementary null result; the system architecture, per-challenge analysis, and live-competition evaluation reported here do not appear there. We will release our code, challenge set, per-run outcome data, and analysis scripts upon publication to support replication.\footnote{\url{https://github.com/jawh3/striatum-ctf}.}

\section{Related Work}

\noindent\textbf{Test-time compute and tool-augmented generation.}
Modern agents scale test-time compute, allocating computation to iterative reasoning during inference~\cite{snell2025scaling}. Early approaches such as Toolformer~\cite{schick2023toolformer} and Gorilla~\cite{patil2024gorilla} baked tool use into the model via fine-tuning, a strategy that becomes brittle as APIs change. Purely neural reasoning strategies such as Chain-of-Thought~\cite{wei2022chain} still suffer from hallucination, confidently emitting syntactically invalid commands. We instead treat the protocol interface as a hard constraint layer, analogous to grammar-constrained decoding~\cite{geng2023grammar,raspanti2025grammar}, grounding the model's reasoning in verified system feedback.

\noindent\textbf{Long-horizon reasoning.}
Methods such as ReAct~\cite{yao2022react} interleave reasoning traces with actions but rely on the context window to hold the full interaction history. In long-horizon settings such as CTF competitions, attention to early-state constraints degrades as the window fills~\cite{liu2024lost}, and a single hallucinated path or misread port scan at step $t$ becomes ground truth at step $t{+}1$, cascading into error propagation~\cite{wang2024survey}. By treating the environment, rather than the token window, as the primary source of truth, STRIATUM-CTF keeps the model focused on the immediate tactical decision.

\noindent\textbf{Automated offensive-security and CTF agents.}
LLM deployment in offensive security has largely followed a ``copilot'' paradigm. PentestGPT~\cite{deng2024pentestgpt} decomposes objectives with a Pentesting Task Tree but keeps a human in the loop to execute generated commands and paste output back, introducing a latency bottleneck that throttles the reasoning loop. Concurrent toolkits such as PentestMCP~\cite{ezetta2025pentestmcp} expose agent-based penetration-testing tools. For CTF specifically, EnIGMA~\cite{abramovich2025enigma} introduces interactive agent--computer interfaces and reaches state-of-the-art results on static benchmarks such as NYU CTF Bench~\cite{shao2024nyuctf} and InterCode-CTF~\cite{yang2023language}, while teams of agents have been shown to exploit real vulnerabilities~\cite{fang2024teams}. Concurrently with our work, the CAI agent won a mixed AI-and-human CTF event using a specialized model architecture~\cite{mayoral2025cai}. STRIATUM-CTF differs from these systems in three ways: it is built entirely from a commodity general-purpose model and off-the-shelf MCP tool servers rather than a specialized model; it replaces the human verifier with deterministic schema validation at the MCP layer, enabling machine-speed iteration; and it is evaluated not only on static benchmarks but in a live, ephemeral competition whose field was composed entirely of human teams, which reduces---though a single event does not eliminate---the contamination risk inherent to published benchmarks (Section~\ref{sec:limits}).

\begin{figure*}[ht]
    \centering
    \includegraphics[width=0.62\linewidth]{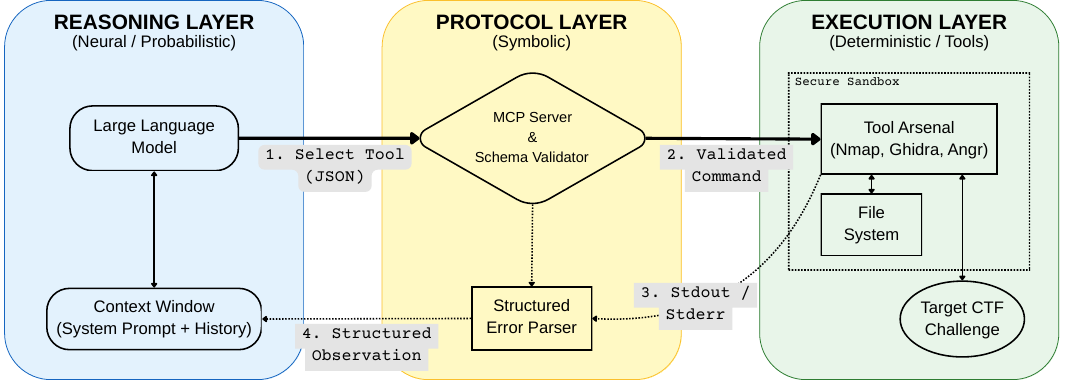}
    \caption{High-level architecture. The system places a validating Protocol Layer between probabilistic reasoning and deterministic execution. The Reasoning Layer (left) emits JSON tool calls that pass through the Protocol Layer (center), which enforces schema validation and rejects malformed calls before invoking tools in the Execution Layer (right). System feedback (stdout/stderr) is structurally parsed and re-injected into the context window, grounding the agent's state in verifiable observations.}
    \label{fig:architecture}
\end{figure*}

\section{Methodology}
\label{sec:method}

We treat CTF solving as sequential decision-making under partial observability: the agent cannot directly observe the target's internal state (e.g., memory layout, open ports) and must act on structured observations returned after each tool invocation. The central design choice is to constrain the agent's action space to schema-valid tool calls, so that malformed actions are rejected before they reach the environment.

\subsection{Architecture}
As shown in Fig.~\ref{fig:architecture}, the system is a closed-loop controller with three layers.
\textbf{The Reasoning Layer} uses Claude Sonnet 4.5 as the planner, with a system prompt establishing its role as an autonomous operator. It produces high-level strategies but cannot execute commands directly.
\textbf{The Protocol Layer} implements MCP as a schema validator, enforcing strict typing on every tool call (e.g., requiring a port number to be an integer in $[1,65535]$). A syntactically invalid call is rejected before it reaches the sandbox.
\textbf{The Execution Layer} consists of security tools (Angr, Ghidra, GDB, and reconnaissance utilities) wrapped in MCP servers, executing in a sandboxed virtual machine and returning structured JSON observations.

\subsection{Operational Definitions}
We use \emph{hallucination} to denote a tool invocation that is syntactically malformed or references a non-existent tool, option, or parameter; such calls are exactly what the Protocol Layer rejects before execution. We distinguish this from a \emph{functional error}, where a well-formed action is semantically wrong (e.g., an incorrect return-address offset); functional errors reach the environment and must be corrected from feedback.

\subsection{The MCP Toolbox}
Alongside a general-purpose command-execution tool, we expose the core analysis primitives as strictly typed, schema-validated functions that enforce type safety and reduce token overhead. The Protocol Layer's constraints bite on these typed functions; the command tool is minimally constrained, and isolating the typed interface's contribution over a raw shell is the comparison we outline in Section~\ref{sec:future}. For \textbf{symbolic and static analysis}, we integrate Angr and Ghidra through atomic functions such as \texttt{solve\_path(addr, cond)} and \texttt{get\_xrefs(func\_name)}; the Protocol Layer enforces input constraints (e.g., valid hexadecimal addresses) and returns either an \texttt{UNSAT} verdict or the satisfying stdin input, abstracting away simulation-state management. For \textbf{dynamic analysis}, a GDB wrapper provides \texttt{inspect\_heap(count)}, transforming raw memory dumps into structured JSON. \textbf{Reconnaissance} operations (via Nmap) are gated by allow-lists restricting execution to in-scope IPv4 ranges and pre-approved binaries.
  
\subsection{Control Loop}
The agent follows a recursive four-step loop. It \emph{plans}, decomposing the objective into a queue of atomic tasks that are dynamically re-ordered as observations arrive. It \emph{executes}, emitting a JSON payload that the Protocol Layer either rejects---a ``circuit breaker'' against malformed parameters---or forwards. It \emph{parses} the returned JSON to extract salient fields (open ports, gadget offsets, memory leaks) while discarding irrelevant data to conserve context. It then \emph{iterates}, evaluating the prior action and refining subsequent steps, or terminating when the objective is met.

\section{Experimental Setup and Results}
\subsection{Dataset and Implementation}
\label{sec:setup}
We assembled a benchmark of 15 CTF challenges from multiple sources: picoCTF\footnote{\url{https://picoctf.org}.} (5 challenges), the official angr.io tutorials\footnote{\url{https://github.com/angr/angr-doc}.} (3 challenges), an archived university CTF (1 challenge), and legacy challenges from internal repositories (6 challenges) for which complete provenance metadata is unavailable. Table~\ref{tab:dataset} summarizes the distribution across categories. The set is intentionally heterogeneous, spanning distinct reasoning chains (e.g., heap traversal vs.\ side-channel timing); its modest size ($N=15$) reflects the manual effort of ground-truth analysis and is a stated limitation (Section~\ref{sec:limits}).

\begin{table}[ht]
    \centering
    \caption{Benchmark dataset composition by category.}
    \label{tab:dataset}
    \begin{tabular}{lc}
        \toprule
        \textbf{Category} & \textbf{Challenges} \\
        \midrule
        Memory Corruption & 4 \\
        Reverse Engineering & 4 \\
        Web Exploitation & 4 \\
        Cryptography & 3 \\
        \midrule
        \textbf{Total} & 15 \\
        \bottomrule
    \end{tabular}
\end{table}

All experiments used Claude Sonnet 4.5 with extended thinking enabled. No sampling parameters were set by the operator; the host's defaults applied, and we are not aware of published values for them. We record the backbone as the model was identified to us, since no build identifier was retained. Experiments ran on an Ubuntu 22.04 LTS virtual machine (16~GB RAM, 8 CPU cores) hosted on an Intel i9-14900HX system. The MCP ecosystem combined community and custom servers: a GDB server (mcp-gdb v0.1.1~\cite{mcp_gdb}), a Ghidra server (GhidraMCP v1.4~\cite{ghidra_mcp} with Ghidra 11.4.2), a custom Angr server wrapping angr 9.2.176, a SecOps server (secops-mcp~\cite{secops_mcp}), a custom command-execution server, and the standard MCP filesystem server. We used the community Claude Desktop for Linux\footnote{\url{https://github.com/aaddrick/claude-desktop-linux}.} (build 0.13.37) as the host.

\subsection{Documentation Ablation}
To test how much performance depends on attached procedural documentation---files placed in Claude Desktop's project workspace and loaded directly into the model's context rather than retrieved (no retrieval-augmented generation)---we evaluated four conditions of increasing richness while holding the model and tool layer fixed. The conditions span from minimal tool schemas to comprehensive guidance (line and token counts in Table~\ref{tab:results}).
\textbf{Minimal} (55 lines) provides only MCP server definitions and tool schemas, and serves as the control.
\textbf{Lessons} (1{,}478 lines) adds lessons distilled from prior debugging sessions plus an automated triage classifier, but no structured templates.
\textbf{Templates} (1{,}976 lines) provides a structured, phase-based solving protocol with attack templates indexed by vulnerability type, plus triage.
\textbf{Baseline} (4{,}147 lines), the system's original full configuration (hence the name), is the union of Lessons and Templates plus orchestration and navigation documents.
Each challenge was attempted three times per condition, yielding $15 \times 4 \times 3 = 180$ runs.

\subsubsection{Evaluation protocol}
Each run was capped at 60 minutes within a single conversation, the condition's documentation loaded once at the start. At roughly ten-minute intervals the operator interrupted the model and sent the token ``Continue'' to resume work in the same conversation---nothing was reset or re-loaded, and operator input never included hints, solution steps, or task-specific reasoning. A run counted as successful only if the agent produced a working exploit recovering the correct flag within the limit, regardless of intermediate failures; reading the flag without a reproducing exploit did not count. A few runs were invalidated and restarted when the agent read a prior run's workspace; e.g., one Templates run on \texttt{sqli\_revenge} was restarted after it opened a previous run's notes.

\subsection{Results}
\label{sec:results}
Table~\ref{tab:results} reports per-condition success rates with Wilson 95\% confidence intervals; Table~\ref{tab:perchallenge} gives the per-challenge outcome matrix. Success rate increases monotonically with documentation, from 77.8\% (Minimal) to 86.7\% (Baseline), but the intervals overlap substantially and the differences are not significant: a $\chi^2$ test of independence on the $4\times2$ outcome table returns $\chi^2(3, N{=}180)=1.36$, $p=0.71$, and a Kruskal--Wallis test on the solve-duration distributions (timed-out runs at the 60-minute cap) returns $p=0.77$. Pairwise effect sizes, measured with Cohen's $h$ (the appropriate statistic for proportions), are small in every case, the largest being Baseline vs.\ Minimal at $h=0.23$. Because trials are correlated within challenges, a generalized-estimating-equations logistic model with challenge-level clustering likewise finds no condition effect (Wald $\chi^2(3){=}3.77$, $p{=}0.29$). These tests indicate absence of evidence for a difference, not demonstrated equivalence.

\begin{table}[htbp]
    \centering
    \caption{Performance across documentation conditions (3 trials per challenge per condition; 180 runs). CI is the 95\% Wilson-score interval. Mean Time is the mean wall-clock minutes per run, with timed-out (unsolved) runs counted at the 60-minute cap; it therefore tracks success rate rather than per-solve speed.}
    \label{tab:results}
    \small
    \setlength{\tabcolsep}{3pt}
    \begin{tabular}{lrrccc}
        \toprule
        \textbf{Condition} & \textbf{Lines} & \textbf{Tokens} & \textbf{Success} & \textbf{95\% CI} & \textbf{Time} \\
        \midrule
        Minimal   & 55      & 591      & 77.8\% (35/45) & 63.7--87.5 & 20.1 \\
        Lessons   & 1{,}478 & 12{,}865 & 82.2\% (37/45) & 68.7--90.7 & 19.1 \\
        Templates & 1{,}976 & 17{,}253 & 84.4\% (38/45) & 71.2--92.3 & 18.5 \\
        Baseline  & 4{,}147 & 36{,}001 & 86.7\% (39/45) & 73.8--93.7 & 17.1 \\
        \bottomrule
    \end{tabular}
\end{table}

\begin{table}[htbp]
    \centering
    \caption{Per-challenge successes (out of 3 trials) under each condition: Minimal (Min), Lessons (Les), Templates (Tem), Baseline (Bas). Categories: M = memory corruption, R = reverse engineering, W = web exploitation, C = cryptography.}
    \label{tab:perchallenge}
    \footnotesize
    \setlength{\tabcolsep}{4pt}
    \renewcommand{\arraystretch}{0.95}
    \begin{tabular}{llcccc}
        \toprule
        \textbf{Challenge} & \textbf{Cat.} & \textbf{Min} & \textbf{Les} & \textbf{Tem} & \textbf{Bas} \\
        \midrule
        \texttt{sample7}            & M & 0 & 0 & 0 & 3 \\
        \texttt{sample3}            & M & 3 & 3 & 3 & 3 \\
        \texttt{sample5}            & M & 3 & 3 & 3 & 3 \\
        \texttt{format\_string\_3}  & M & 2 & 3 & 3 & 3 \\
        \texttt{sample2}            & R & 0 & 0 & 2 & 1 \\
        \texttt{angrs\_00}          & R & 3 & 3 & 3 & 3 \\
        \texttt{angrs\_01}          & R & 3 & 3 & 3 & 3 \\
        \texttt{sample6}            & R & 3 & 3 & 3 & 3 \\
        \texttt{sample8}            & W & 3 & 3 & 3 & 3 \\
        \texttt{sqli\_revenge}      & W & 2 & 2 & 2 & 2 \\
        \texttt{SSTI2}              & W & 1 & 2 & 2 & 2 \\
        \texttt{input\_injection\_1}& W & 3 & 3 & 3 & 3 \\
        \texttt{sample1}            & C & 3 & 3 & 3 & 3 \\
        \texttt{angrs\_02}          & C & 3 & 3 & 3 & 3 \\
        \texttt{power\_analysis\_2} & C & 3 & 3 & 2 & 1 \\
        \midrule
        \textbf{Total (of 45)}      &   & \textbf{35} & \textbf{37} & \textbf{38} & \textbf{39} \\
        \bottomrule
    \end{tabular}
\end{table}

The most striking result is that even the Minimal condition, with only 55 lines of tool schemas and no procedural documentation, reaches 77.8\%. The system is thus robust to a roughly $60\times$ reduction in attached documentation tokens. Because all four conditions share the same MCP-grounded tool layer, this study isolates the contribution of \emph{documentation volume}, not of the protocol layer itself; we discuss this distinction, and the controlled non-MCP comparison it motivates, in Sections~\ref{sec:limits} and~\ref{sec:future}.

The per-challenge matrix (Table~\ref{tab:perchallenge}) sharpens the aggregate picture in three ways. First, nine of the fifteen challenges were solved on every trial under every condition; all between-condition differences concentrate in the remaining six. Restricting a post-hoc analysis of the same 180 runs to those six non-saturated challenges---which removes the ceiling that compresses the aggregate---leaves the ordering intact and the null unchanged: Minimal 44.4\% (8/18), Lessons 55.6\% (10/18), Templates 61.1\% (11/18), Baseline 66.7\% (12/18), with $\chi^2(3, N{=}72)=1.98$, $p=0.58$. The Baseline--Minimal gap does widen to 22.2~pp ($h=0.45$) once the saturated challenges are removed, so this result is properly read as ``not distinguishable at this sample size'' rather than as a demonstrated small effect: with 18 trials per condition the study is underpowered against differences of that magnitude. Second, most of the aggregate spread comes from a single memory-corruption challenge (\texttt{sample7}) that only the full-documentation condition solved. Third, the differences are not uniformly in Baseline's favor: on \texttt{power\_analysis\_2} the Minimal condition succeeded 3/3 while Baseline succeeded 1/3, and on \texttt{sample2} Templates (2/3) outperformed Baseline (1/3). Durations tell a consistent story: among successful runs only, Minimal had the \emph{lowest} mean solve time (8.7 min vs.\ 10.5 min for Baseline), so the aggregate mean-time advantage of richer documentation comes entirely from fewer timeouts, not faster solves. We therefore caution against reading the monotonic trend as evidence that more documentation reliably helps: operator notes recorded the opposite effect on some runs---under the richest documentation the agent occasionally ``forgets things and ignores instructions''---context drift, in which the volume of attached guidance crowds out the agent's attention to its own state~\cite{liu2024lost}.

\subsubsection{Case studies}
Two challenges illustrate the mechanism. On the timing side-channel challenge \texttt{sample2}, which requires exploiting execution-time variance to recover a PIN, the agent without timing-attack guidance defaulted to symbolic execution, which is ineffective for side channels; only the conditions containing the structured attack templates recovered the PIN at all (2/3 and 1/3, Table~\ref{tab:perchallenge}). In contrast, \texttt{angrs\_00} was solved in three minutes or less under every condition, because it maps directly onto the Angr server's \texttt{solve\_by\_output} primitive that even the Minimal schema exposes. The value of documentation is thus challenge-dependent, largest for vulnerability classes whose effective approach the tool primitives do not directly suggest. This pattern is consistent with the companion study's hypothesis that high-bandwidth, schema-validated environment feedback substitutes for procedural guidance, leaving documentation valuable mainly where feedback alone cannot reveal the right approach~\cite{chacko2026skills}.

\subsubsection{Schema-mediated error recovery}
We do not measure a hallucination rate quantitatively, because the per-call execution logs from the original runs were not retained (Section~\ref{sec:limits}); the qualitative structure, however, is visible in our records. Because the Protocol Layer rejects malformed tool calls before execution, a class of errors that would otherwise corrupt downstream state is converted into an immediate, structured rejection the agent can act on by revising the call. Operator notes corroborate self-correction on functional errors as well: on one memory-corruption run the agent recovered from repeated tool-internal timeouts and solved the challenge after 27 minutes once it ``fixed itself.'' We can, however, name one recurring failure mode we observed that the Protocol Layer does not guard against: \emph{hangs}---a well-formed, schema-valid call that the server accepted and then never returned from, consuming the run's remaining time budget. Schema validation constrains the \emph{form} of a call, not the \emph{liveness} of the server that answers it, so this class of failure passes through the Protocol Layer untouched. We did not instrument failure modes and therefore report no frequencies, and we do not claim this mode was the most common. A practical corollary for anyone building on this design is that per-call timeouts and server health checks belong alongside schema validation. We present these as qualitative observations; a quantitative comparison of schema-validated versus raw-shell tool interfaces is the most important open measurement, outlined in Section~\ref{sec:future}.

\subsection{Live CTF Competition}
\begin{figure}[htbp]
    \centering
    \includegraphics[width=0.95\linewidth]{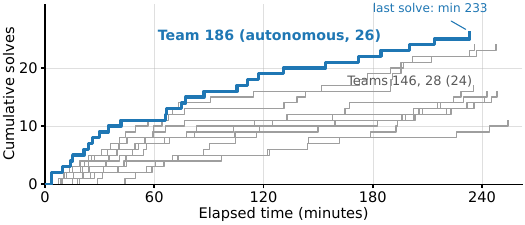}
    \caption{Cumulative challenge solves of the top 10 teams. The autonomous system (Team~186, highlighted) recorded its final solve at minute 233 and ended with the most solves (26 vs.\ 24); the gray curves are the human teams. Final ranking was determined by points, reported in the text.}
    \label{fig:ctf_top_10}
\end{figure}

To evaluate the system outside a controlled benchmark, a team using STRIATUM-CTF competed in the Florida State University (FSU) CTF, a live multi-team competition held on November 15, 2025. The event was open to undergraduate and graduate students, most of whom had taken FSU's ``Offensive Computer Security'' course; challenges emphasized binary exploitation and web security. More than 30 teams registered and 22 (including ours) solved at least one challenge. Fig.~\ref{fig:ctf_top_10} shows the cumulative-solve progression of the top 10 teams over the four-hour event.

STRIATUM-CTF finished first with 215 points, ahead of the strongest human teams at 205 (second) and 200 (third); the remaining teams scored below 135. We state the scope of this result before interpreting it: it is a single, non-replicated event with a 15-point margin separating first from third, and run outcomes were not deterministic---identical conditions produced different outcomes on five of the fifteen challenges in the controlled study---so it demonstrates feasibility rather than statistical superiority over human experts. For replicability we record the live-run configuration. The agent ran on the same six-server MCP stack as the controlled study (verified against a configuration snapshot taken five days after the event), under the same Claude Desktop for Linux build noted above, with extended thinking enabled, and the same backbone and unset sampling parameters noted in Section~\ref{sec:setup}. It was deployed with the \textbf{Baseline} documentation set unmodified. We note the chronology: the competition preceded the controlled study by several weeks, so the deployed configuration reflects the system as it then stood rather than a choice informed by the ablation. Together with a concurrent win by a specialized agent in a mixed AI-and-human event~\cite{mayoral2025cai}, this adds evidence that autonomous agents can now win live CTFs outright; distinctively, our result was achieved by a commodity general-purpose model with off-the-shelf MCP tooling, in a field composed entirely of human teams. The agent's edge came from operating at machine speed---tool calls and output parsing complete without the latency of manual command entry---and from tireless, systematic retrying of failed exploits.

\section{Discussion}

\subsection{Threats to Validity}
\label{sec:limits}
We state the scope of our claims explicitly. The controlled study establishes that an MCP-grounded agent solves a large majority of challenges and that this performance is robust to large reductions in attached documentation. Because all four conditions share the same tool layer, the study bounds the effect of documentation volume rather than isolating the protocol layer itself; we therefore treat schema-mediated grounding as a design principle---motivated by the tool-use failure modes discussed earlier and supported by the qualitative error-recovery behavior reported above---whose quantitative causal effect is the object of the schema-versus-shell comparison in Section~\ref{sec:future}. Three factors bound generalization. First, the benchmark draws partly on public picoCTF and angr-tutorial material that may overlap the model's training data and is modest in size (several challenges succeed under every condition), so it characterizes tool integration rather than contamination-free generalization; the live competition complements it on exactly this axis, as its challenges were authored for the event and not previously published and the agent met them only at the start (the authors are associated with the host course but did not write the challenges). Second, the evaluation uses a single backbone (Claude Sonnet 4.5) and reports tool reliability qualitatively. The per-call execution logs were not retained because the runs were conducted interactively in a desktop host that does not persist per-call MCP traces, and no logging proxy was interposed between the client and the tool servers; the instrumentation was an omission in the original design rather than a deliberate exclusion, and Section~\ref{sec:future} names the mechanism that closes it. Third, the backbone used in both the controlled study and the live competition is no longer served, so these specific numbers cannot be reproduced exactly on the original model. The artifacts we release therefore support replication of the \emph{protocol} on current backbones rather than re-execution of the original runs: the tool stack and challenge set are fully specified above, but the backbone itself cannot be reconstituted. We accordingly report the mechanism qualitatively rather than attach a rejection rate obtained from a different model under a different host, which would not describe the results presented here; Section~\ref{sec:future} specifies the instrumented design that would supply that number properly. A further caveat follows from the hang failure mode described above: because an unreturned tool call consumes a run's time budget, some recorded failures reflect tool-server availability rather than the agent's reasoning. This adds variance that is unrelated to the documentation manipulation, and is a further reason to read the null result as absence of evidence rather than as demonstrated equivalence.

\subsection{Implications for Defense and Education}
If results like ours and CAI's~\cite{mayoral2025cai} replicate, the consequences for defenders and educators are immediate. For defenders, autonomous offense at machine speed means detection and response practices calibrated to human attacker tempo are no longer conservative---and the building blocks of such offense (a commodity model, public MCP tool servers) are freely available. For educators, an open-entry CTF can no longer be assumed to measure individual skill: an agent assembled from public components placed first in a field of student teams. The same systems, however, offer scalable sparring partners for offensive-security courses and always-available red-team automation. Course assessments should shift toward proctored or process-oriented evaluation, and defensive curricula should incorporate exercises against agentic adversaries.

\subsection{Future Work}
\label{sec:future}
These caveats define a concrete program. The decisive next experiment would be a $2\times K$ crossed design that holds the challenge set fixed while crossing the tool interface---schema-validated MCP vs.\ raw shell, with the same binaries installed in both arms so that capability is held constant and only the typed interface varies---against $K$ backbones spanning a current frontier proprietary model and an open-weights model of comparable scale, which simultaneously tests model-invariance. Both arms would be instrumented by a stdio proxy interposed between the client and each MCP server, tee-ing every call and its response to a per-run log. That proxy is precisely the component whose absence produced the reporting gap in Section~\ref{sec:limits}, and it supplies the four metrics the present study cannot: schema-rejection rate, invalid-call rate, tool-call success fraction, and error-recovery latency. Because the primary endpoint is measured per tool call rather than per run, such a design yields substantially more analyzable events per challenge, though the clustering that motivates the model in Section~\ref{sec:results} would still apply. Further natural extensions include benchmarking on community suites such as NYU CTF Bench~\cite{shao2024nyuctf} for direct comparison with prior agents, and reducing MCP's token overhead through hierarchical context management. We will release the code, challenge set, per-run outcome data, and analysis scripts upon publication to support these extensions.

\section{Conclusion}
We presented STRIATUM-CTF, an agentic framework that mediates between LLM reasoning and tool execution through the Model Context Protocol, rejecting malformed tool calls before execution and grounding the agent in structured feedback. In a 180-run controlled study the system succeeded on up to 86.7\% of attempts and remained robust to a roughly $60\times$ reduction in attached documentation, with no statistically significant difference between conditions; in a single live competition it placed first against human teams. We report these results with explicit scope: protocol-mediated grounding is a practical mechanism for filtering malformed tool calls and enabling machine-speed iteration, but isolating its causal contribution requires the schema-versus-shell comparison we outline as future work. STRIATUM-CTF provides an early, reproducible reference point for protocol-grounded autonomous offensive-security agents.

\section*{Ethical Considerations and Responsible Use}
All experiments were conducted in authorized, sandboxed settings. Benchmark challenges were public CTF artifacts or instructor-authorized course material, and the live competition was a sanctioned, opt-in event in which our team was a registered participant. The agent's reconnaissance tools are gated by IPv4 allow-lists and pre-approved-binary lists (Section~\ref{sec:method}) that confine activity to in-scope targets. We acknowledge the dual-use nature of autonomous offensive tooling: capabilities that accelerate defensive red-teaming and security education can also lower the barrier to misuse. We therefore release no exploit payloads or attack-template corpora and follow coordinated disclosure for any novel vulnerability surfaced; quantifying machine-speed offensive capability helps defenders and educators calibrate accordingly.

\section*{AI Tools Used}
The system under study uses Anthropic's Claude Sonnet 4.5 as its reasoning engine; this model is the object of study, not an author. Separately, in preparing this manuscript the authors used Anthropic's Claude for assistance throughout: drafting, revising, editing, and formatting author-written text, and writing and cross-checking the analysis scripts, whose outputs the authors verified against the raw run data. No generative tool produced experimental results, and all research design, experimentation, analysis, claims, and conclusions are the authors' own. The authors take full responsibility for the entire content of this paper, consistent with IEEE policy.

\bibliographystyle{ieeetr}
\bibliography{ijcnn26}

\begin{thebibliography}{10}

\bibitem{aggarwal2024discovering}
P.~Aggarwal, S.~Venkatesan, J.~Youzwak, R.~Chadha, and C.~Gonzalez,
  ``Discovering cognitive biases in cyber attackers' network exploitation
  activities: A case study,'' in {\em Human Factors in Cybersecurity, 15th
  International Conference on Applied Human Factors and Ergonomics (AHFE
  2024)}, 2024.

\bibitem{vaswani2017attention}
A.~Vaswani, N.~Shazeer, N.~Parmar, J.~Uszkoreit, L.~Jones, A.~N. Gomez,
  {\L}.~Kaiser, and I.~Polosukhin, ``Attention is all you need,'' in {\em
  Proceedings of the 31st International Conference on Neural Information
  Processing Systems}, pp.~6000--6010, 2017.

\bibitem{hakim2025neuro}
S.~B. Hakim, M.~Adil, A.~Velasquez, S.~Xu, and H.~H. Song, ``Neuro-symbolic
  {AI} for cybersecurity: State of the art, challenges, and opportunities,''
  {\em arXiv preprint arXiv:2509.06921}, 2025.

\bibitem{happe2025surprising}
A.~Happe and J.~Cito, ``On the surprising efficacy of {LLMs} for
  penetration-testing,'' {\em arXiv preprint arXiv:2507.00829}, 2025.

\bibitem{stechly2025self}
K.~Stechly, K.~Valmeekam, and S.~Kambhampati, ``On the self-verification
  limitations of large language models on reasoning and planning tasks,'' in
  {\em The Thirteenth International Conference on Learning Representations},
  2025.

\bibitem{park2025slopsquatting}
S.~Park, ``Slopsquatting: Hallucination in coding agents and vibe coding,''
  {\em Trend Micro}, 2025.

\bibitem{liu2024lost}
N.~F. Liu, K.~Lin, J.~Hewitt, A.~Paranjape, M.~Bevilacqua, F.~Petroni, and
  P.~Liang, ``Lost in the middle: How language models use long contexts,'' {\em
  Transactions of the Association for Computational Linguistics}, vol.~12,
  pp.~157--173, 2024.

\bibitem{sood2025paradigm}
A.~K. Sood, S.~Zeadally, and E.~Hong, ``The paradigm of hallucinations in
  {AI}-driven cybersecurity systems: Understanding taxonomy, classification
  outcomes, and mitigations,'' {\em Computers and Electrical Engineering},
  vol.~124, p.~110307, 2025.

\bibitem{mcp_docs}
{Anthropic}, ``{Model Context Protocol}.''
  \url{https://modelcontextprotocol.io}, 2025.

\bibitem{shoshitaishvili2016sok}
Y.~Shoshitaishvili, R.~Wang, C.~Salls, N.~Stephens, M.~Polino, A.~Dutcher,
  J.~Grosen, S.~Feng, C.~Hauser, C.~Kruegel, {\em et~al.}, ``{SoK}: ({State
  of}) {The Art of War}: Offensive techniques in binary analysis,'' in {\em
  2016 IEEE symposium on security and privacy (SP)}, pp.~138--157, IEEE, 2016.

\bibitem{ghidra2019}
{National Security Agency}, ``Ghidra: Software reverse engineering framework.''
  \url{https://github.com/NationalSecurityAgency/ghidra}, 2019.

\bibitem{chacko2026skills}
S.~J. Chacko, J.~Hugglestone, C.~M. Islam, and X.~Liu, ``When skills don't
  help: A negative result on procedural knowledge for tool-grounded agents in
  offensive cybersecurity,'' in {\em Proceedings of the First Workshop on Agent
  Skills, co-located with ACM Conference on AI and Systems (CAIS)}, 2026.
\newblock arXiv:2605.20023.

\bibitem{snell2025scaling}
C.~V. Snell, J.~Lee, K.~Xu, and A.~Kumar, ``Scaling {LLM} test-time compute
  optimally can be more effective than scaling parameters for reasoning,'' in
  {\em The Thirteenth International Conference on Learning Representations},
  2025.

\bibitem{schick2023toolformer}
T.~Schick, J.~Dwivedi-Yu, R.~Dess{\`\i}, R.~Raileanu, M.~Lomeli, E.~Hambro,
  L.~Zettlemoyer, N.~Cancedda, and T.~Scialom, ``Toolformer: Language models
  can teach themselves to use tools,'' {\em Advances in Neural Information
  Processing Systems}, vol.~36, pp.~68539--68551, 2023.

\bibitem{patil2024gorilla}
S.~G. Patil, T.~Zhang, X.~Wang, and J.~E. Gonzalez, ``Gorilla: Large language
  model connected with massive {APIs},'' {\em Advances in Neural Information
  Processing Systems}, vol.~37, pp.~126544--126565, 2024.

\bibitem{wei2022chain}
J.~Wei, X.~Wang, D.~Schuurmans, M.~Bosma, B.~Ichter, F.~Xia, E.~H. Chi, Q.~V.
  Le, and D.~Zhou, ``Chain-of-thought prompting elicits reasoning in large
  language models,'' in {\em Proceedings of the 36th International Conference
  on Neural Information Processing Systems}, pp.~24824--24837, 2022.

\bibitem{geng2023grammar}
S.~Geng, M.~Josifoski, M.~Peyrard, and R.~West, ``Grammar-constrained decoding
  for structured {NLP} tasks without finetuning,'' in {\em Proceedings of the
  2023 Conference on Empirical Methods in Natural Language Processing},
  pp.~10932--10952, 2023.

\bibitem{raspanti2025grammar}
F.~Raspanti, T.~Ozcelebi, and M.~Holenderski, ``Grammar-constrained decoding
  makes large language models better logical parsers,'' in {\em Proceedings of
  the 63rd Annual Meeting of the Association for Computational Linguistics
  (Volume 6: Industry Track)}, pp.~485--499, 2025.

\bibitem{yao2022react}
S.~Yao, J.~Zhao, D.~Yu, N.~Du, I.~Shafran, K.~R. Narasimhan, and Y.~Cao,
  ``{ReAct}: Synergizing reasoning and acting in language models,'' in {\em The
  Eleventh International Conference on Learning Representations}, 2023.

\bibitem{wang2024survey}
L.~Wang, C.~Ma, X.~Feng, Z.~Zhang, H.~Yang, J.~Zhang, Z.~Chen, J.~Tang,
  X.~Chen, Y.~Lin, {\em et~al.}, ``A survey on large language model based
  autonomous agents,'' {\em Frontiers of Computer Science}, vol.~18, no.~6,
  p.~186345, 2024.

\bibitem{deng2024pentestgpt}
G.~Deng, Y.~Liu, V.~M. Vilches, P.~Liu, Y.~Li, Y.~Xu, M.~Pinzger, S.~Rass,
  T.~Zhang, and Y.~Liu, ``{PentestGPT}: Evaluating and harnessing large
  language models for automated penetration testing,'' in {\em USENIX Security
  Symposium}, 2024.

\bibitem{ezetta2025pentestmcp}
Z.~Ezetta and W.-c. Feng, ``{PentestMCP}: A toolkit for agentic penetration
  testing,'' {\em arXiv preprint arXiv:2510.03610}, 2025.

\bibitem{abramovich2025enigma}
T.~Abramovich, M.~Udeshi, M.~Shao, {\em et~al.}, ``{EnIGMA}: Interactive tools
  substantially assist language model agents in finding security
  vulnerabilities,'' in {\em Proceedings of the 42nd International Conference
  on Machine Learning (ICML)}, 2025.

\bibitem{shao2024nyuctf}
M.~Shao, S.~Jancheska, M.~Udeshi, {\em et~al.}, ``{NYU CTF Bench}: A scalable
  open-source benchmark dataset for evaluating {LLMs} in offensive security,''
  in {\em Advances in Neural Information Processing Systems (NeurIPS), Datasets
  and Benchmarks Track}, 2024.

\bibitem{yang2023language}
J.~Yang, A.~Prabhakar, S.~Yao, K.~Pei, and K.~R. Narasimhan, ``Language agents
  as hackers: Evaluating cybersecurity skills with capture the flag,'' in {\em
  Multi-Agent Security Workshop@ NeurIPS'23}, 2023.

\bibitem{fang2024teams}
Y.~Zhu, A.~Kellermann, A.~Gupta, {\em et~al.}, ``Teams of {LLM} agents can
  exploit zero-day vulnerabilities,'' {\em arXiv preprint arXiv:2406.01637},
  2024.

\bibitem{mayoral2025cai}
V.~Mayoral-Vilches, L.~J. Navarrete-Lozano, F.~Balassone, {\em et~al.},
  ``Cybersecurity {AI}: The world's top {AI} agent for security
  capture-the-flag ({CTF}),'' {\em arXiv preprint arXiv:2512.02654}, 2025.

\bibitem{mcp_gdb}
{Signal Slot Inc.}, ``{MCP GDB Server for Debugging with AI Assistants}.''
  \url{https://github.com/signal-slot/mcp-gdb}, 2025.

\bibitem{ghidra_mcp}
{Maintainers of ghidraMCP}, ``{MCP Server for Ghidra}.''
  \url{https://github.com/LaurieWired/GhidraMCP}, 2025.

\bibitem{secops_mcp}
{SecurityforTech}, ``{Security Operations Multi-Tool Platform (MCP)}.''
  \url{https://github.com/securityfortech/secops-mcp}, 2025.

\end{thebibliography}

\end{document}